\documentclass[argument]{aastex631}

\usepackage{hyperref}
\usepackage{amsmath}
\usepackage{enumerate}
\usepackage[utf8]{inputenc}
\usepackage{cleveref}
\usepackage{amssymb}
\usepackage[T1]{fontenc}
\usepackage{enumitem}
\usepackage{url}
\usepackage[misc]{ifsym}
\usepackage{multirow}
\usepackage{CJK}
\graphicspath{{./}{figures/}}

\begin{document}
\begin{CJK*}{UTF8}{gbsn}
\title{Polarization Criterion in Targeted SETI Observation}
\shortauthors{Li et al.}

\author[0000-0002-1190-473X]{Jian-Kang Li}
\affiliation{Institute for Frontiers in Astronomy and Astrophysics, Beijing Normal University, Beijing 102206, China}
\affiliation{Department of Astronomy, Beijing Normal University, Beijing 100875, China;\url{tjzhang@bnu.edu.cn}}
\affiliation{Department of Astronomy, Yunnan University, Kunming 650500, China}

\author{Yu Chen}
\affiliation{Institute for Frontiers in Astronomy and Astrophysics, Beijing Normal University, Beijing 102206, China}
\affiliation{Department of Astronomy, Beijing Normal University, Beijing 100875, China;\url{tjzhang@bnu.edu.cn}}

\author[0000-0002-8719-3137]{Bo-Lun Huang}
\affiliation{Institute for Frontiers in Astronomy and Astrophysics, Beijing Normal University, Beijing 102206, China}
\affiliation{Department of Astronomy, Beijing Normal University, Beijing 100875, China;\url{tjzhang@bnu.edu.cn}}

\author[0000-0002-4683-5500]{Zhen-Zhao Tao}
\affiliation{Institute for Frontiers in Astronomy and Astrophysics, Beijing Normal University, Beijing 102206, China}
\affiliation{Department of Astronomy, Beijing Normal University, Beijing 100875, China;\url{tjzhang@bnu.edu.cn}}

\author[0000-0003-3977-4276]{Xiao-Hang Luan}
\affiliation{Institute for Frontiers in Astronomy and Astrophysics, Beijing Normal University, Beijing 102206, China}
\affiliation{Department of Astronomy, Beijing Normal University, Beijing 100875, China;\url{tjzhang@bnu.edu.cn}}

\author[0000-0002-3464-5128]{Xiao-Hui Sun}
\affiliation{Department of Astronomy, Yunnan University, Kunming 650500, China}

\author[0000-0002-3363-9965]{Tong-Jie Zhang（张同杰）\href{mailto:tjzhang@bnu.edu.cn}{\textrm{\Letter}}}
\affiliation{Institute for Frontiers in Astronomy and Astrophysics, Beijing Normal University, Beijing 102206, China}
\affiliation{Department of Astronomy, Beijing Normal University, Beijing 100875, China;\url{tjzhang@bnu.edu.cn}}

\author[0000-0002-8604-106X]{Vishal Gajjar}
\affiliation{Breakthrough Listen, University of California Berkeley, Berkeley, California, 94720, USA}
\affiliation{SETI Institute, 339 N Bernardo Ave Suite 200, Mountain View, CA 94043, USA}



\begin{abstract}
In this paper, we propose a novel method for distinguishing extraterrestrial intelligence (ETI) signals from radio frequency interference (RFI) by leveraging polarization features. We exploit the sinusoidal variation of the linearly polarized components of Stokes parameters with the parallactic angle as a characteristic signature of ETI signals, while such linearly polarized components remain relatively stable for terrestrial RFI. Typically, a minimum of 4-8 hr of observation time is required to detect these sinusoidal variations. The polarization approach in the search for extraterrestrial intelligence also enables us to study the radio stellar bursts emitted by M-type stars as ancillary science, which is relevant to assessing the habitability of exoplanets. Compared to the frequency drift method, the polarization method effectively reduces the required observation time for signal identification while improving the signal identification process.

\end{abstract}
\keywords{\href{http://astrothesaurus.org/uat/2127}{Search for extraterrestrial intelligence(2127)}; \href{http://astrothesaurus.org/uat/1278}{Polarimetry(1278)}; \href{http://astrothesaurus.org/uat/2128}{Technosignatures(2128)}}


\section{Introduction}

\par In the field of radio search for extraterrestrial intelligence (SETI), there are generally two types of observation, namely sky survey and targeted observation \citep{Tarter2001TheSF}. The sky surveys have been conducted since the early development of SETI, such as SERENDIP, SETI@home \citep{Werthimer2001BerkeleyRA} and Five-hundred-meter Aperture Spherical radio Telescope (FAST) SETI backend \citep{Zhang2020FirstSO}. In recent years, targeted SETI observations have gained popularity among prominent projects like the Breakthrough Listen (BL) initiative (e. g. \citealt{Enriquez2017TheBL,Price2020TheBL,Sheikh2020TheBL,Gajjar2021TheBL}), Allen Telescope Array (ATA) \citep{Harp2016SETIOO,Harp2020AnAS} and FAST targeted SETI observation \citep{Tao2022SensitiveMT,Luan2023MultibeamBS}. However, most of these observations focus solely on the total received intensity rather than the variation of each polarization component. Nowadays, many single-dish radio telescopes and telescope arrays are equipped with orthogonal dual-polarization receivers, which enables us to add polarization measurement as a new parameter for signal identifications.

\par It is reasonable to infer that extraterrestrial intelligence (ETI) signals may exhibit certain polarization features since our own technologies produce polarized radio emissions, many of which are used for communication. Leveraging these potential features, we propose a novel method for identifying ETI signals based on their polarization characteristics. Specifically, we explore the variation of the Stokes parameters, which are used to quantify polarization, with the parallactic angle resulting from the Earth's rotation. Additionally, polarization in SETI observations allows us to investigate radio stellar bursts from M-type stars, which has implications for space weather and the habitability of exoplanets.

\section{Polarization Features of the Received Signals}\label{sec:polarfea}
\par Generally speaking, there are three types of signals that could be received by radio telescopes: ETI signals, natural astrophysical emissions, and radio frequency interference (RFI). Each of these types may exhibit specific polarization features. For instance, commonly observed natural astrophysical radio sources such as pulsars, masers, and stars can exhibit certain polarization \citep{tinbergen_1996}. Statistical analysis of data from the FAST telescope has also revealed that terrestrial RFI signals are predominantly polarized \citep{Zhang2021RadioFI}. As most polarized astrophysical radiations in nature are broadband signals, they can be easily distinguished by their frequency bandwidth. However, our primary concern lies in differentiating between ETI signals and RFI. ETI signals are generally assumed to be narrowband, as most naturally occurring signals are known to be wideband  ($>$ 1 kHz; \citealt{Cohen1987NarrowPC}), whereas many RFIs can also be narrowband signals. RFIs can be further classified into terrestrial RFI and satellite RFI. To distinguish ETI signals from RFI, we propose utilizing variations in the Stokes parameters for signal identification.
\par
\begin{figure}
\centering
\includegraphics[scale=0.65]{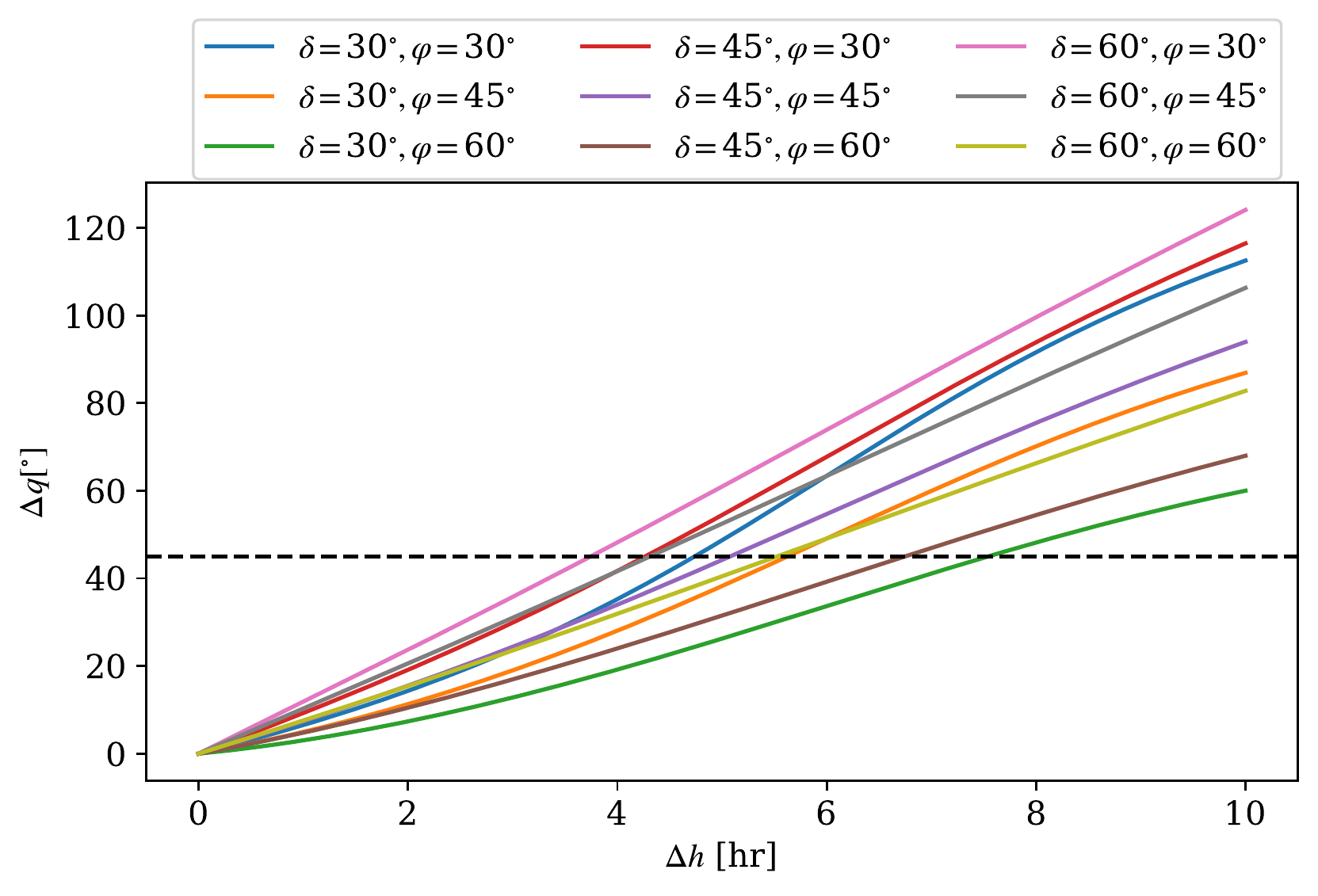}
\caption{\label{handq}The change of parallactic angle with hour angel for observation targets at different declinations and telescopes at different latitudes, respectively. The black horizontal dashed line represents the case that $\Delta q=45^{\circ}$.}
\end{figure}

 The all-Stokes observation can be measured by a telescope with two orthogonal linear polarization X and Y, which can produce autocorrelation data XX and YY along with cross-correlation data XY and YX via Fourier transform within the integration time. The Stokes vector $\mathbf{S}$ can be written by
   \begin{equation}
   \mathbf{S}=
   \left[ \begin{array}{c}
   	I\\
   	Q\\
   	U\\
   	V
   \end{array} \right]=
   \left[ \begin{array}{c}
   	XX+YY\\
   	XX-YY\\
   	2XY\\
   	-2YX
   \end{array} \right]
   \label{Stokesobs}
   \end{equation} 
    where $I$ represents the total intensity, $Q$ and $U$ are the linearly polarized components and $V$ is the circularly polarized component. In real observation, the field of view with respect to the observer should rotate with Earth, and the Stokes parameters we receive should be 
  \begin{subequations}
\begin{eqnarray}
I'&=&I, \label{StokestelescopeI}\\ 
Q'&=&Q\cos 2q+U\sin 2q, \label{StokestelescopeQ}\\ 
U'&=&-Q\sin2q+U\cos2q ,\label{StokestelescopeU}\\
V'&=&V,\label{StokestelescopeV}
\end{eqnarray}
  \end{subequations}
where $q$ is the parallactic angle defined as the angle between the great circle through a celestial object and the zenith, and the hour circle of the object, which can be calculated by \citep{avila1991field}
\begin{equation}
\tan q=\frac {\sin h}{\cos \delta \tan \varphi -\sin \delta \cos h}.
\end{equation}
where $h$ is the local hour angle, $\delta$ is the decl. of the source, and $\varphi$ is the latitude of the telescope. Equation (\ref{StokestelescopeQ}) and (\ref{StokestelescopeU}) can be rewritten into a uniform sine or cosine form by the auxiliary angle formula (See Appendix), and there is a phase difference of 45$^{\circ}$ between $Q'$ and $U'$. Different decl. of targets and latitudes of telescopes can cause different changes in parallactic angles; therefore, whether the phase difference can be witnessed also depends on $\delta$ and $\varphi$. Typically, telescopes located at higher latitudes and observing targets at lower declinations require longer observation times compared to those located at lower latitudes and targeting objects at higher declinations (See Fig \ref{handq}). 
To witness such a phase difference of 45$^{\circ}$ between $Q'$ and $U'$, at least 4-8 hr of observation time are required and recommended. 
For terrestrial RFIs, their Stokes parameters $Q'$ and $U'$ should not change with $q$, since terrestrial RFI sources are kept stationary to the telescope. In targeted SETI observation, the satellite RFIs can be detected when the satellite is orbiting exactly in the observation area and disappears when the satellite orbits out of the observation area. In SETI observations, in addition to natural astrophysical sources, the received signal at a given frequency may consist of three contributors: system noise, RFI, and potential ETI signals. The received signals in the $X$ and $Y$ directions are separate and independent. Assuming that the noise intensity can be modeled by a complex normal distribution in both $X$ and $Y$ directions and at all observed frequencies, the resulting four correlation data should follow a chi-squared distribution after Fourier transform \citep{Brzycki2020NarrowbandSL}. As an example, we generate synthetic observation data for linearly polarized ETI and RFI by combining the chi-squared noise. The noise intensity is calculated based on data from targeted observations of Kepler-438 using the FAST telescope. Notably, an event of interest was identified in the FAST observation of Kepler-438 around 1140.6040 MHz \citep{Tao2022SensitiveMT}.
\par
\begin{figure*}
\centering
\includegraphics[scale=0.5]{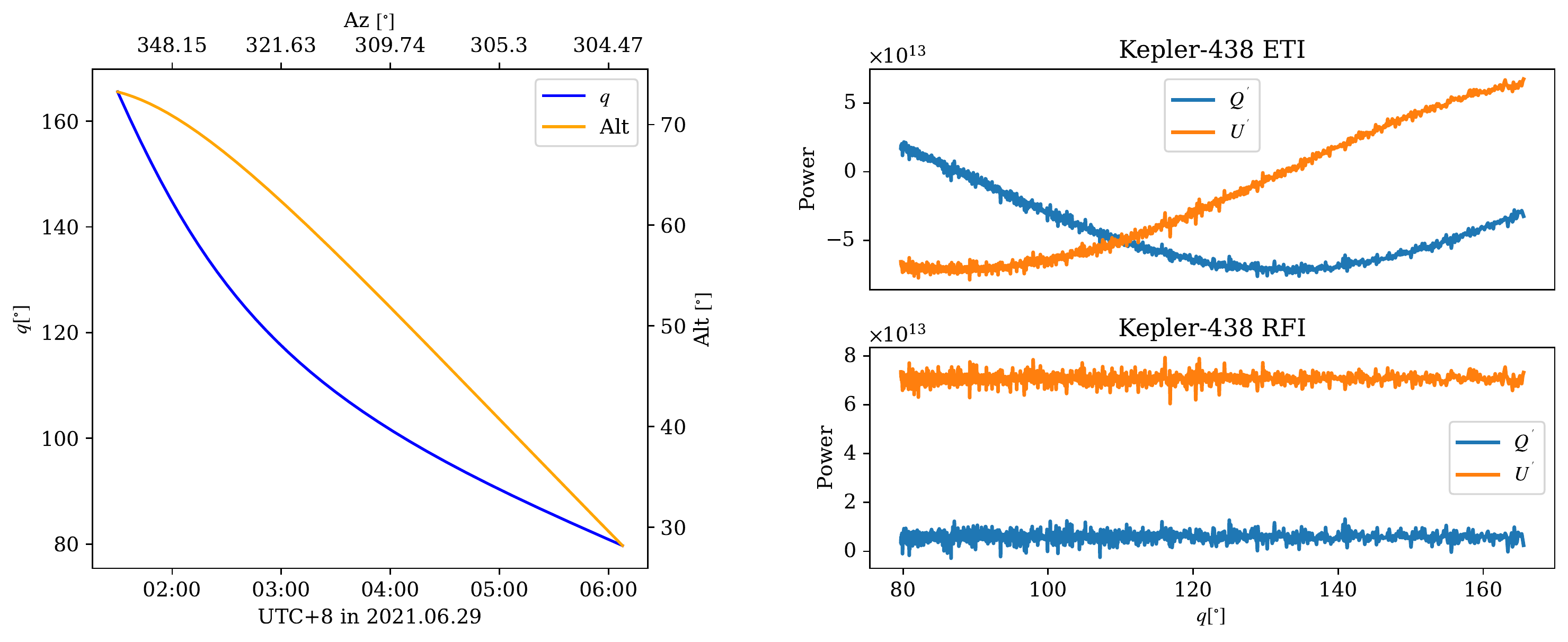}
\caption{\label{QandU}Left: the changes of parallactic angle, altitude, and azimuth of Kepler-438 during the $\sim$4 hr observation in local time 2021.06.29 of FAST. Right top: the variation of  $Q'$ and $U'$ of the synthetic ETI signal during the $\sim$4 hr observation of Kepler-438. Right bottom: the variation of  $Q'$ and $U'$ of the synthetic RFI during the $\sim$4 hr observation of Kepler-438.}
\end{figure*}

 In order to witness the phase difference between $Q'$ and $U'$, the observation should be long enough so that the change of parallactic angle can be greater than 45$^{\circ}$. We set the observation time to $\sim$4 hr for ETI and terrestrial RFI synthetic observation.
 4 hr observation time is sufficient for the change in $q$ greater than 45$^{\circ}$ (see the left panel of Fig \ref{QandU}). The variations and the phase difference between $Q'$ and $U'$ are obvious in the waterfall plot, while the Stokes parameters of RFI are relatively stable. Since the waterfall plot only presents general polarization features of the synthetic observation data, we extract the $Q'$ and $U'$ data of synthetic ETI signal and synthetic RFI to further analyze the polarization features quantificationally (see the two right panels of Fig \ref{QandU}). The small fluctuation is caused by the chi-squared background noise. $Q'$ and $U'$ of synthetic RFI only fluctuate around the mean values, while the synthetic ETI signal shows sinusoidal variations and 45$^{\circ}$ phase difference between $Q'$ and $U'$.
 
\section{Comparison with Frequency Drift}
\subsection{Observation Time}
\par Frequency drifting is the most commonly used method in SETI. ETI signals are expected to appear Doppler frequency drift due to the relative motion between transmitter and receiver. For the commonly discussed Earth-exoplanet case, the relative motion is mainly contributed from the rotations and orbits of Earth and exoplanet. The frequency drift can be quantified by drift rate, which is an important parameter and is used to search possible ETI narrowband signal and distinguish RFI in SETI research.  Since the orbits of planet are much longer than the observation time, the frequency drift of the signal are usually regarded as linear change with constant drift rate.
\par TurboSETI \citep{Enriquez2017TheBL,2019ascl.soft06006E} is the python package to search narrowband drifting signal with high-spectral resolution in frequency.  Most of the previous targeted observations only focus Stokes-$I$ data with certain appropriate drift rates, while many RFIs can also appear similar frequency drift property. Although many observational and computational methods have been applied in removing RFI from the observation data, there is still no method that can separates all RFIs from the observation data. A solution is to carry out long-term observation for a certain source. If the observation time is long enough, an ETI signal should exhibit a pseudosinusoidal frequency variation due to the asymmetric orbital motions of Earth and exoplanet \citep{Li2022DriftRO}. Since the orbital periods of the discovered exoplanets are, according to the statistic data from exoplanet archive, \footnote{\url{https://exoplanetarchive.ipac.caltech.edu/}} most of them are smaller than the Earth's orbital period, the frequency drift is usually contributed by the orbits of exoplanets. For habitable planets, they can not be too close or too far from the its host star, which makes their orbital periods neither too short nor too long. Such rotational periods are usually in the range of ten days to a few months. To witness the peak or trough of the pseudosinusoidal frequency variation is more practicable for real observation, nevertheless, this would still requires a quarter of the exoplanets' orbital periods. The polarization method requires less observation time, as we discuss in Section \ref{sec:polarfea}, since the sinusoidal variations of Stokes-$Q$ and Stokes-$U$ only depend on the rotation of Earth, which is relatively shorter than the orbital periods of exoplanets. And the phase difference between Stokes-$Q$ and Stokes-$U$, the more useful criterion in polarization method, can be witnessed for several hours, which is more practicable for limited observation time.

\subsection{Sensitivity}
\par The minimum detectable flux can be given by \cite{Enriquez2017TheBL}, \cite{Gajjar2021TheBL},\cite{2021NatAs...5.1148S} and \cite{Tao2022SensitiveMT}
\begin{equation}
    f_{\min}=\frac{\text{S/N}_{\min}}{\beta}\frac{2k_BT_{\text{sys}}}{A_{\text{eff}}}\sqrt{\frac{\delta\nu}{n_{\text{pol}}\tau_{\text{obs}}}},
\end{equation}
where $\text{S/N}_{\min}$ is the signal-to-noise threshold, $\beta$ is the dechirping efficiency, $k_B$ is the Boltzmann constant, $T_{\text{sys}}$ is the system temperature, $A_{\text{eff}}$ is the effective collecting area, $\delta\nu$ is the frequency resolution, $n_{\text{pol}}$ is the polarization number, and $\tau_{\text{obs}}$ is the observation time. For high-spectral-resolution data, signals with high drift rates can spread to across multiple frequency channels, and the dechirping efficiency can be expressed as \citep{Gajjar2021TheBL} 
\begin{equation}
    \beta =\begin{cases}
	1&		0<\left| \dot{\nu} \right|\le \dot{\nu}_{\min}\\
	\frac{\dot{\nu}_{\min}}{\left| \dot{\nu} \right|}&		\dot{\nu}_{\min}<\left| \dot{\nu} \right|\le \dot{\nu}_{\max},\\
\end{cases}
\end{equation}
where $\dot{\nu}$ is the drift rate of the signal, $\dot{\nu}_{\min}$ is the minimum drift rate defined as 
\begin{equation}
    \dot{\nu}_{\min}=\frac{\delta\nu}{\tau_{\text{obs}}}
\end{equation}
and $\dot{\nu}_{\max}$ is the maximum drift rate defined as 
\begin{equation}
    \dot{\nu}_{\max}=\frac{B_{\text{obs}}}{\delta t}
\end{equation}
with $B_{\text{obs}}$ being the total bandwidth of the observation and $\delta t$ being the time resolution \citep{2019ApJ...884...14S}. For the traditional search for narrowband drifting signal in Stokes-$I$ data, only XX and YY polarization channels are used, so $n_{\text{pol}}$ is 2. For the search in \citep{Tao2022SensitiveMT}, signals are searched in XX and YY polarization channels separately, so $n_{\text{pol}}$ is 1. Although this method can increase the possibility to find more candidates, on the other hand, the minimum detectable flux becomes $\sqrt{2}$ times the former one. The polarization method in this work can increase the sensitivity of the observation since all the four polarization channels' data are used and $n_{\text{pol}}$ is 4 and $f_{\min}$ becomes $1/\sqrt{2}$ of the one corresponding to the search in total intensity data.

\section{Stellar Burst in SETI Observation and Polarimetry}\label{sec:stellarburst}
\par Although our primary interest lies in detecting ETI signals emitted from exoplanets, our current technology does not yet allow us to directly focus our telescopes on individual exoplanets during real observations. Instead, what we observe is the planetary system consisting of a star and the planets orbiting it. In addition to the potential ETI signals from exoplanets, it is highly likely that we can detect some interesting radiation emissions from the host stars themselves. M-type stars, the most common type of stars in the Milky Way \citep{Henry2006TheSN}, have been the focus of numerous studies on the search for Earth-like planets and their habitability  \citep{Shields2016TheHO,Childs2022LifeOE,doAmaral2022TheCO}. Due to their strong magnetic fields (typically kilogauss; \citealt{JohnsKrull1995DetectionOS,Kochukhov2020MagneticFO}), longer rotational braking times \citep{Delfosse1998RotationAC,Barnes2003OnTR,Delorme2011StellarRI}, and higher activity levels for a given rotation period \citep{Kiraga2007AgeRotationActivityRF}, the proportion of active stars among M-type stars is much larger than among G-type stars, and the flare activities, a generic results of magnetic reconnection \citep{Pettersen1989ARO}, occur more frequently and energetically in M-type stars than in solar flares \citep{Lacy1976UVCS,Haisch1989AnOO,Odert2008HOM,Davenport2016THEKC}. 
\par Stellar radio flares typically last from seconds to hours and cover frequencies ranging from megahertz to gigahertz. Observations over the past decades have revealed that stellar radio bursts are commonly characterized by close to 100\% circular polarization \citep{Osten2006WideBandSO,Osten2008UltrahighTR,Zic2019ASKAPDO,Zic2020AFI} with extremely high brightness temperatures exceeding  $10^{14}$ K \citep{Gdel1989BroadbandSO,Bastian1990DynamicSO}. Such circularly polarized bursts can be visualized by Stokes-$I$ and Stokes-$V$ waterfall plots (usually with time as the horizontal axis and frequency as the vertical axis) for the full observation band. Many of these stellar flares also exhibit frequency drift properties due to the intrinsic motion, and the drift rate can be written as \citep{2002ASSL..279.....B}
\begin{equation}
\dot{\nu}\approx -\nu \frac{v_s\cos \phi}{2H_n}
,
\label{driftratestellarsim}
\end{equation} 
where $\nu$ is frequency, $\phi$ angle between the beam direction and the vertical, $v_s$ is the propagation velocity and $H_n$ is the density scale height. For given observed drift rate, frequency, and scale height (which can be estimated from X-ray observations), we can make constrain on the propagation velocity, or we can obtain the estimation for scale height $H_n$, which is relevant to the hydrostatic equilibrium of the star, from the observed drift rate, frequency, and the constraint on propagation velocity. 
\par Stellar flares are not only relevant to space weather but also have implications for the habitability of exoplanets. Planets located in the habitable zone around M-type stars are likely to be tidally locked \citep{Barnes2017TidalLO}, and since M-type stars have lower luminosities, the habitable zone is closer to the star. As a result, the atmospheres of such planets are exposed to a more severe magnetic environment, facing challenges such as magnetospheric compression and atmospheric erosion \citep{Khodachenko2007CoronalME,Lammer2007CoronalME}, which ultimately affect their habitability. Kepler-438 b, a potential habitable planet, is likely experiencing energetic stellar flares from its host star occurring approximately every 200 days \citep{Armstrong2016TheHS}. Given the low probability of detecting signals of interest in SETI observations and the limited resources of telescopes, including stellar radio bursts as a scientific goal in SETI observations not only enhance the efficiency of data utilization but also enable the study of star magnetic activity and space weather through obtained observations. These data can be utilized to assess the habitability of exoplanets and optimize the selection of future observational targets.

\section{Conclusion}\label{sec:conclusion}
\par We present polarization as a novel and effective physical criterion in SETI observations, which identifies the potential ETI signal by the sinusoidal variation of Stokes parameters with the parallactic angle of the source. This method can be applied to any radio telescope equipped with two orthogonal linear polarization directions. 
\par Firstly, compared to the commonly used physical criterion of frequency drift, polarization reduces the observation time required to identify ETI signals. Frequency drift in signals emitted from transmitters on exoplanets is caused by the rotations and orbits of the exoplanet and Earth, with the exoplanet's orbit typically contributing more significantly due to the relatively shorter orbital periods of habitable planets around M-type stars. This frequency drift can be characterized by a pseudosinusoidal curve of the frequency change \citep{Li2022DriftRO}. To observe such a curve, a relatively long observation time is needed since the orbital periods of exoplanets are still much longer than our observation time. For instance, Kepler-438 b has an orbital period of 35.23319 days \citep{Torres2015VALIDATIONO1}, Even to capture the peak or trough portions of the curve, at least a quarter of the exoplanet's orbital period is required. As discussed in previous sections, several hours of observation time are sufficient to detect the sinusoidal variations in $Q'$ and $U'$, as the change in parallactic angle depends solely on the rotation of Earth for a given observation location and target.
\par Secondly, polarization allows us to include stellar bursts as a scientific goal in targeted SETI observations, significantly enhancing the utilization of observational data. Most observed radio stellar bursts exhibit high circular polarization, and the polarization features of these bursts can be characterized and visualized using Stokes-$I$ and Stokes-$V$ waterfalls for intensity and circular polarization, respectively. Studies on stellar bursts provide valuable insights into the magnetic activity of stars, space weather, and the habitability of exoplanets. Assessing habitability can further optimize the selection of future target observations.
\par Thirdly, compared with the traditional method of search in total intensity data in the previous observations or the method of search in XX and YY polarization channels separately, the all-Stokes polarization observation can increase the sensitivity of the observation, since the polarization method can make full use of the data of four polarization channels and decrease the minimum detectable flux. Based on Sections \ref{sec:polarfea} - \ref{sec:stellarburst}, we can conclude the following:
\begin{enumerate}
\item The sinusoidal variations of Stokes parameters can be applied to the identification of ETI signal, while the Stokes parameters of terrestrial RFI should be relatively stable.
\item The sinusoidal variations of Stokes parameters for ETI signal identification require at least several hours to visualize in order that we can witness the phase difference of the sine curves.
\item Polarization enables us to include stellar bursts as a scientific goal in SETI observation since most of the radio stellar bursts are highly circularly polarized. And studies on radio stellar bursts can offer us information about space weather and the habitability of exoplanets.
\item Polarization can effectively reduce the observation time required for signal identification compared to frequency drift methods. 
\item All-Stokes polarization observation can increase the sensitivity of the observation compared with the traditional methods.
\end{enumerate}

\vspace{\baselineskip}
We thank the anonymous referees for the the insightful and useful comments that helped us greatly improve this article. T.-J.Z (张同杰) dedicates this paper to the memory of his mother, Yu Zhen Han (韩玉珍) who passed away 3 yr ago (2020 August 26). This work was supported by the National Science Foundation of China (grant No. 11929301) and the National SKA Program of China (2022SKA0110202).



%

\appendix\label{app:auxiliary}
\section*{Uniform form of $Q'$ and $U'$}
Let 
\begin{equation}
    I_L=\sqrt{Q'^2+U'^2}
\end{equation}
be the intensity of the linearly polarized component. Using the auxiliary angle formula, Equation (\ref{StokestelescopeQ}) and (\ref{StokestelescopeU}) can be rewritten in to
\begin{subequations}
\begin{eqnarray}
Q'&=&I_L\sin(2q+\arctan\frac{Q}{U}), \label{StokestelescopeQuni}\\ 
U'&=&I_L\sin(2q-\arctan\frac{U}{Q})\nonumber\\
&=&I_L\sin2(q-\chi).\label{StokestelescopeUuni}
\end{eqnarray}
  \end{subequations}
while
\begin{equation}
    \arctan\frac{Q}{U}+\arctan\frac{U}{Q}=\frac{\pi}{2}.
\end{equation}
Then, Equation (\ref{StokestelescopeQuni}) can be rewritten as 
\begin{equation}
    Q'=I_L\sin[2(q-\chi)+\frac{\pi}{4}].
\end{equation}

\bibliography{sample631}
\bibliographystyle{aasjournal}



\end{CJK*}
\end{document}